\documentclass[twocolumn,showpacs,amsfonts,aps,prc,nofootinbib,floatfix,%
superscriptaddress]{revtex4}

\usepackage{amsmath}
\usepackage{bm}
\usepackage{graphicx}

\usepackage{epsfig}
\newcommand{\beq}{\begin{equation}}
\newcommand{\eeq}{\end{equation}}
\newcommand{\bea}{\vspace{0.25cm}\begin{eqnarray}}
\newcommand{\eea}{\end{eqnarray}}

\newcommand{\ro}{\mbox{{\boldmath
$\rho$}}}

\newcommand{\db}{{{\bf d}}}
\newcommand{\rb}{\mbox{{\bf
r}}}

\newcommand{\bb}{{{\bf b}}}

\newcommand{\E}{{{\bf E}}}

\newcommand{\Vb}{{{\bf V}}}

\newcommand{\B}{{{\bf B}}}

\def\lsim{\mathrel{\rlap{\lower4pt\hbox{\hskip1pt$\sim$}}
    \raise1pt\hbox{$<$}}}         
\def\gsim{\mathrel{\rlap{\lower4pt\hbox{\hskip1pt$\sim$}}
    \raise1pt\hbox{$>$}}}         

\newcommand{\landau}{L.D.~Landau Institute for Theoretical Physics,
        GSP-1, 117940, Kosygina Str. 2, 117334 Moscow, Russia}

\begin{document}


\title{
Effect of giant resonances on fluctuations of 
electromagnetic fields in heavy ion collisions
}
\date{\today}

\author{B.G.~Zakharov}\affiliation{\landau}

\begin{abstract}
We perform quantum calculations of fluctuations of the electromagnetic 
fields in $AA$ collisions at RHIC and LHC energies. 
Calculations are performed with the help of the 
fluctuation-dissipation theorem accounting for the giant dipole and quadrupole 
resonances. We find that in the quantum picture the field fluctuations 
are much smaller than that predicted by the classical Monte-Carlo simulation 
with the Woods-Saxon nuclear density used in previous analyses.

\end{abstract}
%

\maketitle

\section{Introduction}
A very strong  magnetic 
can be generated in heavy ion collisions at RHIC and LHC energies:
 $eB\sim 3m_{\pi}^{2}$ for RHIC ($\sqrt{s}=0.2$ TeV)
and $eB\sim 45m_{\pi}^{2}$ for LHC ($\sqrt{s}=2.76$ TeV)
\cite{Kharzeev_B1,Toneev_B1,Tuchin_B,Z_maxw}.
In the last years effect of the magnetic field 
on the processes in the quark-gluon plasma (QGP) produced 
in $AA$ collisions attracted much attention 
(e.g.,  the charge separation along the magnetic field direction
due to the anomalous current $\propto \B$ 
(the chiral magnetic effect) \cite{Kharzeev_B1,Kharzeev_CME_rev},
the synchrotron photon emission \cite{T1,Z_syn}, anisotropy in the 
heavy quark diffusion \cite{HQ_dif1,HQ_dif2}, the magnetohydrodynamic 
flow effects \cite{MHD1,MHD2,MHD3}). 
To a good approximation \cite{Z_maxw}, 
the initial fields
after intersection of the Lorentz contracted nuclei
are determined by a sum of the fields generated by the colliding nuclei.
But at later times the QGP response can modify them \cite{Skokov_B,Z_maxw}. 
If one neglects it,
the average electric, $\langle \E\rangle$, and magnetic, $\langle \B\rangle$,
fields of each nucleus are simply given by the Lorentz transformation of 
its Coulomb field in the
nucleus rest frame. The total average magnetic field, 
in the center of mass system of the nucleus-nucleus collision, 
at $y=0$ (here $y$ 
is the axis transversal to the reaction plane, as shown in Fig.~1) turns out 
to be transversal to the reaction plane. However, the field fluctuations
can destroy this picture.
For study of the medium electromagnetic effects in $AA$ collisions
it is important to know magnitude of the field fluctuations.
\begin{figure} [t]
\vspace{.7cm}
\begin{center}
\includegraphics[height=3.5cm]{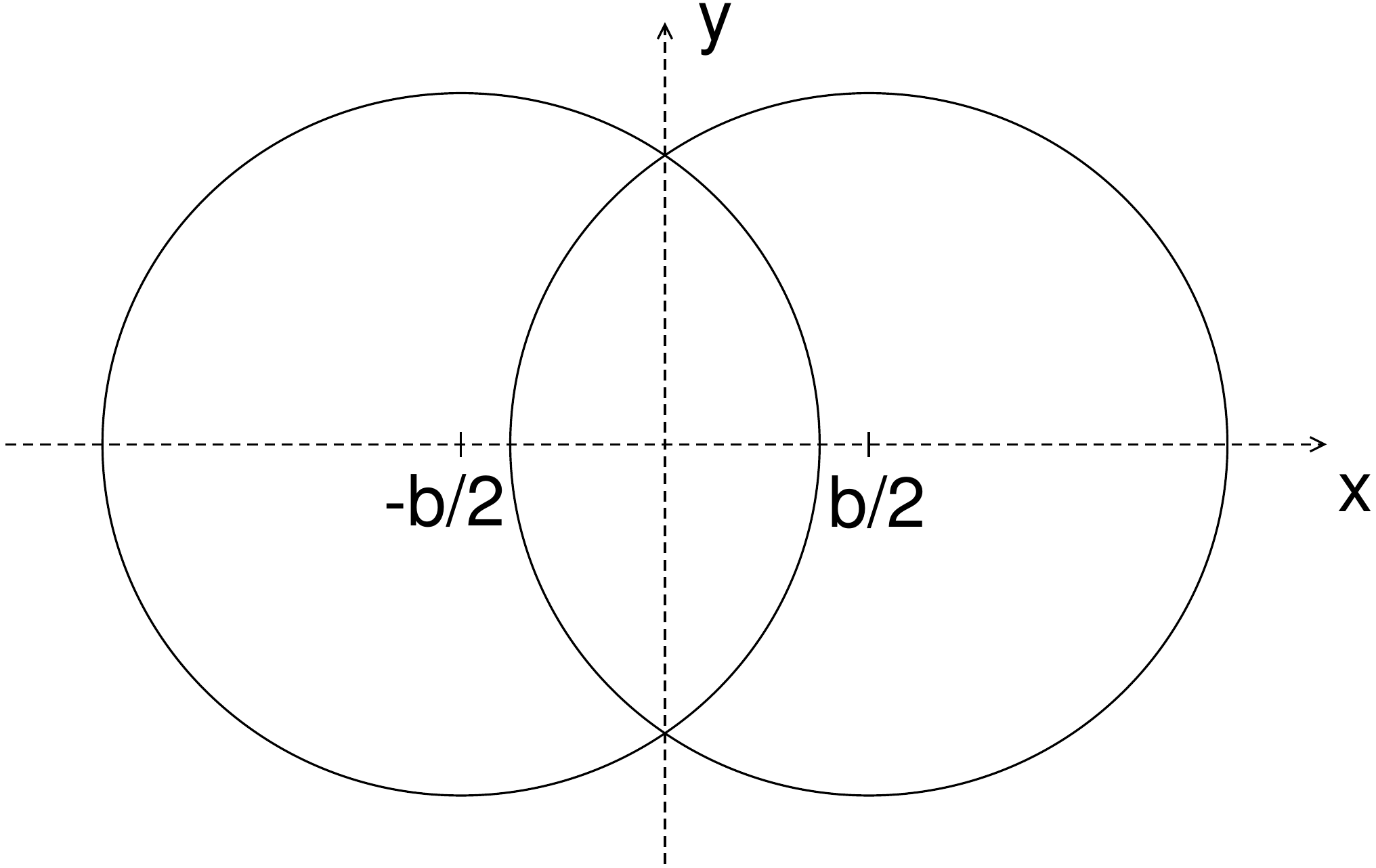}
\end{center}
\vspace{-0.5cm}
\caption[.]
{The transverse plane of a non-central $AA$ collision with the impact
parameter $b$. 
}
\end{figure}
Usually, in the literature 
(see, e.g., Refs.~\cite{Skokov_MC,Deng_MC,Liao_MC,Roy_MC})
fluctuations of the electromagnetic fields
in $AA$ collisions are treated using the classical Lienard-Weichert potentials
of the protons within the Monte-Carlo simulation
with the Woods-Saxon (WS) nuclear distributions.
This approach gives rather large event-by-event fluctuations 
of the magnetic field (both parallel and perpendicular to the reaction plane).
However, the classical Monte-Carlo treatment has no serious theoretical
justification. The deviations from the classical approach may come
both from the dynamical quantum effects in the colliding nuclei
and from the quantum effects for the electromagnetic fields.
Indeed, the field fluctuations should be most sensitive 
to the large scale fluctuation of the electric charge density 
in the colliding nuclei.
It is well known that such large scale fluctuations 
in heavy nuclei are dominated by the collective
giant resonances (see, e.g., Refs.~\cite{BM, Greiner,Speth,Roca}). 
From the point of view of fluctuations of the 
electromagnetic fields in heavy ion collisions the potentially 
important collective excitations are the isovector 
giant dipole resonance (IV-GDR)
and  isoscalar/isovector giant quadrupole resonances (IS/IV-GQRs) 
with the energy $\omega_R\sim 10-25$ MeV \cite{Speth,Roca} 
(the isoscalar modes
correspond to the shape vibrations of the nucleus as a whole, and for the
isovector ones protons and neutrons oscillate out of phase).  
But the factorized WS nuclear distribution ignores the collective 
quantum effects.
From the side of the electromagnetic field the classical
treatment should be invalid when the distance from the nucleus, $R$, 
in the nucleus rest frame,  becomes bigger than $1/\omega_R$.
Since for the central rapidity region 
(i.e. at $z$ close to zero in the center mass frame) 
$R\sim \tau\gamma$ (here $\tau$ is the proper time and $\gamma$ is the Lorentz
factor), it means that the classical model fails
already at the proper time $\tau\gsim 0.1$ fm for RHIC energies
and at $\tau\gsim 0.01$ fm for LHC energies. 

The quantum calculation of the electromagnetic field fluctuations
in $AA$ collisions can be performed using  
the general formulas of the fluctuation-dissipation theorem (FDT) \cite{Callen} 
for the electromagnetic fluctuations given in \cite{LL9}.
In the case of interest the field fluctuations 
can be expressed via the nuclear dipole and quadruple polarizabilities.
The contribution of the dipole mode have been addressed in 
Ref.~\cite{Z1}. The results of this analysis show that
in the quantum picture the field fluctuations turn out to be much smaller 
than predictions of the classical Monte-Carlo simulation with the 
WS nuclear density. It is highly desirable to perform
the quantum calculation including the quadrupole modes. Although the
contribution of the quadrupole modes decease steeper with $R$ 
than that for the dipole mode, they  potentially may become 
important in the region of small proper time (which corresponds to small 
$R$), where the effects of the magnetic fields should be strongest.
In the present letter we address the field fluctuations accounting for
both the GDR and GQRs.

\section{Theoretical framework}
We consider $AA$ collision between right moving and left moving
nuclei  with velocities
$\Vb_{R}=(0,0,V)$ and $\Vb_{L}=(0,0,-V)$, and with the impact parameters
$\bb_R=(0,-b/2)$ and $\bb_L=(0,b/2)$ (as shown in Fig.~1).
We take $z_{R,L}=\pm Vt$.
We evaluate the electromagnetic fields generated by two colliding
nuclei in the ground state. 
As in Ref.~\cite{Z1}, we ignore the electromagnetic fields generated
by the induced currents in the QGP created after $AA$ collision.
The total electromagnetic field is a sum of the fields generated by
the colliding nuclei.
For each nucleus, we write the electromagnetic field 
as a sum of the mean field and the fluctuating field
\beq
F^{\mu\nu}=\langle F^{\mu\nu}\rangle+\delta F^{\mu\nu}\,.
\label{eq:10}
\eeq
$\langle \E\rangle$ and $\langle \B\rangle$  are given 
by the Lorentz transformation of its Coulomb field in the nucleus rest frame.
The mean magnetic field for two colliding nuclei at $\rb=0$  has 
only $y$-component.
Simple calculations give for the total mean $y$-component of the magnetic
field at 
$t^{2}\gsim (R_{A}^{2}-b^{2}/4)/\gamma^{2}$ (here $R_{A}$ is the 
nucleus radius, and $b$ is assumed to be $<2R_{A}$)
\beq
\langle B_{y}(t,\rb=0)\rangle \approx \frac{\gamma Zeb}{(b^{2}/4+\gamma^{2}V^2t^{2})^{3/2}}\,.
\label{eq:20}
\eeq
At $t\gg R_{A}/\gamma$    
in the region $\rho\ll t\gamma$ $\langle B_{y}(t,\ro,z=0)\rangle$      
takes a simple $\rho$-independent form 
\beq
\langle B_{y}(t,\ro,z=0)\rangle \approx
Zeb/\gamma^{2}t^{3}\,.
\label{eq:30}
\eeq

For each colliding nucleus, we first calculate the 
correlators of the fluctuating electromagnetic
fields in the nucleus rest frame, and then perform the Lorentz transformation
to  the center of mass lab-frame of $AA$ collision.
We use the FDT formalism for electromagnetic fluctuations of 
\cite{LL9}, formulated in the gauge $\delta A^{0}=0$. 
It relates the time Fourier component of the vector potential 
correlator
\bea
\langle \delta A_i(\rb_1)\delta A_k(\rb_2)\rangle_{\omega}=
\frac{1}{2}\int dt e^{i\omega t}
\langle \delta A_i(t,\rb_1)\delta A_k(0,\rb_2)
\nonumber\\
+\delta A_k(0,\rb_2)\delta A_i(t,\rb_1)\rangle\hspace{2cm}
\label{eq:40}
\eea
to the retarded Green function
\bea
D_{ik}(\omega,\rb_1,\rb_2)=-i\int dt e^{i\omega t}
\theta(t)
\langle \delta A_i(t,\rb_1)\delta A_k(0,\rb_2)
\nonumber\\
-\delta A_k(0,\rb_2)A_i(t,\rb_1)
\rangle\,.\hspace{2cm}
\label{eq:50}
\eea
For the case of the zero temperature, that we need,  the FDT relation between 
(\ref{eq:40}) 
and (\ref{eq:50}) reads \cite{LL9} 
\bea
\!\!\langle \delta A_i(\rb_1)\delta A_k(\rb_2)\rangle_{\omega}\!=\!
-\mbox{sign}(\omega)
\mbox{Im} D_{ik}(\omega,\rb_1,\rb_2).
\label{eq:60}
\eea

In vacuum the Green function is given by \cite{LL9}
\beq
D_{ik}^{v}(\omega,\rb_1,\rb_2)=\delta_{ik}D_1(\omega,r)+
\frac{x_ix_k}{r^2}D_2(\omega,r)\,,
\label{eq:70}
\eeq
where $\rb=\rb_1-\rb_2$, and
\beq
D_1(\omega,r)=-\frac{e^{i\omega r}}{r}\left(1+\frac{i}{\omega r}
-\frac{1}{\omega^2 r^2}\right)\,,
\label{eq:80}
\eeq
\beq
D_2(\omega,r)=\frac{e^{i\omega r}}{r}\left(1+\frac{3i}{\omega r}
-\frac{3}{\omega^2 r^2}\right)\,.
\label{eq:90}
\eeq
The vacuum Green function corresponds to the ordinary vacuum fluctuations
of electromagnetic fields. For our purpose in this work, we 
need only correction to the vacuum Green function 
at $\rb_{1}=\rb_2$ due to interaction
of electromagnetic fields with the nuclei.
We calculate it assuming that $R=|\rb_{1,2}-\rb_A|$ is large 
as compared to the nucleus radius $R_A$, when the nucleus can be treated
as a point-like object described by the nucleus polarizability
tensors.  
In the present analysis we account for the effect of the dipole and 
quadrupole nucleus modes. 
The dipole contribution to
$\Delta  D_{ik}$ can written as \cite{LL9}
\bea
\Delta  D_{ik}^{d}(\omega,\rb_1,\rb_2)=-
\omega^2 D_{il}^{v}(\omega,\rb_1,\rb_A)\alpha_{lm}(\omega)\nonumber\\
\times D_{mk}^{v}(\omega,\rb_A,\rb_2)\,.
\label{eq:100}
\eea
where $\alpha_{ik}(\omega)$ is the dipole nuclear polarizability tensor.
The tensor $\alpha_{ik}(\omega)$   can be written 
as \cite{LL4,Migdal_GDR}
\beq
\alpha_{ik}(\omega)=\sum_{s}\left[
\frac{\langle 0|\db_i|s\rangle \langle s|\db_k|0\rangle }
{\omega_{s0}-\omega -i\delta}+
\frac{\langle 0|\db_k|s\rangle \langle s|\db_i|0\rangle  }
{\omega_{s0}+\omega +i\delta}\right]\,,
\label{eq:110}
\eeq
where $\db$ is the dipole operator.
The formulas (\ref{eq:100}), (\ref{eq:110}) correspond to the dipole 
approximation
of the electromagnetic interaction operator $V=-\db \E$ \cite{LL4}.
For the quadrupole excitations the electromagnetic interaction operator
reads 
$V=-\frac{1}{6}\sum_{i,j} Q_{ij}\frac{\partial E_i}
{\partial x_j}$ 
\cite{LL2,Greiner_PR151}, 
where $Q_{ij}$ is the operator
of the quadrupole moment.
And the quadrupole counterpart of the dipole contribution to
$\Delta  D_{ik}$ (\ref{eq:100})  can be written as
\bea
\Delta  D_{ik}^{q}(\omega,\rb_1,\rb_2)=-
\omega^2 \frac{\partial}{\partial \rb_A^{j}}D_{il}^{v}(\omega,\rb_1,\rb_A)
\alpha_{ljmn}(\omega)\nonumber\\
\times 
\frac{\partial}{\partial \rb_A^{n}}D_{mk}^{v}(\omega,\rb_A,\rb_2)\,,
\label{eq:120}
\eea
where now $\alpha_{ljmn}(\omega)$ is the quadrupole 
nuclear polarizability tensor. The quadrupole counterpart of the 
representation (\ref{eq:110}) for the dipole tensor 
$\alpha_{ik}(\omega)$ is  given by
\bea
\alpha_{ijmn}(\omega)=\frac{1}{6^2}\sum_{s}\left[
\frac{\langle 0|Q_{ij}|s\rangle \langle s|Q_{mn}|0\rangle }
{\omega_{s0}-\omega -i\delta}\right.
\nonumber\\+
\left.\frac{\langle 0|Q_{mn}|s\rangle \langle s|Q_{ij}|0\rangle  }
{\omega_{s0}+\omega +i\delta}\right]\,.
\label{eq:130}
\eea
The formulas (\ref{eq:100}), (\ref{eq:120})
correspond to the approximation of a point-like nucleus with
the dipole and quadrupole moments.
Note that the applicability condition 
$R/R_A\gg 1$ (here, as above, $R=|\rb_{1,2}-\rb_{A}|$ is the distance from the 
observation point at $\rb_1=\rb_2$ to the center of the nucleus
in its rest frame) for this approximation
means that the time $t$ in the center of mass lab-frame 
of $AA$ collision for the center of the QGP fireball  must be large as compared to $R_A/\gamma$.

The field correlators that we need can be written as 
\beq
\hspace{-.07cm}\langle \delta E_i(t,\rb)\delta E_k(t,\rb)\rangle
=\frac{1}{2\pi}\int_{-\infty}^{\infty}d\omega
\langle \delta E_i(\rb)\delta E_k(\rb)\rangle_{\omega}\,,
\label{eq:140}
\eeq
\beq
\hspace{-.08cm}\langle \delta B_i(t,\rb)\delta B_k(t,\rb)\rangle
=\frac{1}{2\pi}\int_{-\infty}^{\infty}d\omega
\langle \delta B_i(\rb)\delta B_k(\rb)\rangle_{\omega}\,.
\label{eq:150}
\eeq
Here the time Fourier components of 
the electromagnetic field correlators in terms of that for the 
the vector potential correlator (\ref{eq:40}) are given by
\beq
\langle \delta E_i(\rb_1)\delta E_k(\rb_2)\rangle_{\omega}=
\omega^{2}\langle \delta A_i(\rb_1)\delta A_k(\rb_2)\rangle_{\omega}\,,
\label{eq:160}
\eeq
\beq
\hspace{-.18cm}
\langle \delta B_i(\rb_1)\delta B_k(\rb_2)\rangle_{\omega}=
\mbox{rot}^{(1)}_{il}
\mbox{rot}^{(2)}_{kj}
\langle \delta A_l(\rb_1)\delta A_j(\rb_2)\rangle_{\omega},
\label{eq:170}
\eeq
where the vector potential correlator
should be calculated using (\ref{eq:60}) with replacement of the full tensor 
$D_{ik}$ by
the correction $\Delta D_{ik}$ due to interaction of the electromagnetic field
with the dipole (\ref{eq:100}) and quadrupole (\ref{eq:120}) modes. 

We will consider the spherically symmetrical even-even
$^{208}$Pb nucleus. In this case the dipole and quadrupole polarization tensors
can be written in terms of two scalar functions $\alpha_d=\alpha_{ii}$ and
$\alpha_q=\alpha_{ijij}$ as  
\beq
\alpha_{ij}=\alpha_d\delta_{ik}/3\,,
\label{eq:180}
\eeq
\beq
\alpha_{ijkl}=\frac{1}{10}\alpha_q\left[\delta_{ik}\delta_{jl}+\delta_{il}\delta_{jk}-
\frac{2}{3}\delta_{ij}\delta_{kl}\right]\,.
\label{eq:190}
\eeq
The relation (\ref{eq:190}) can be obtained using the fact that the quadrupole
operator satisfies the relation $Q_{ii}=0$. 
It is important that $\alpha_d$ and
$\alpha_q$ are analytical functions of $\omega$ 
in the upper half-plane \cite{LL4}, and satisfy the 
relation $\alpha^{*}(-\omega^{*})=\alpha(\omega)$ \cite{LL4}.
It allows, for both the modes, to transform the integrals over $\omega$ 
in (\ref{eq:140}), (\ref{eq:150})
from $-\infty$ to $\infty$ to those along the positive
imaginary axis, where $\alpha_{d,q}$ are real.

We parametrize the functions $\alpha_{d,q}$ by
a single Lorentzian form
\beq
\alpha_i(\omega)=
c_i 
\left[
\frac{1}{\omega_{i}-\omega -i\Gamma_i/2}+
\frac{1}{\omega_{i}+\omega +i\Gamma_i/2}
\right]\,.
\label{eq:200}
\eeq
The imaginary part of the polarizability is proportional to the
photoabsorption cross section \cite{LL4,Greiner_PR151}. 
The GDR   for $^{208}$Pb due to $E1$ transition  
is well seen in the data on the photoabsorption
cross section \cite{GDR_Pb}. 
In Ref.~\cite{Z1} by fitting the photoabsorption
cross section for 
$^{208}$Pb   from Ref.~\cite{GDR_Pb}  
 we obtained for the dipole mode:
$\omega_{d}\approx 13.3$ MeV, $\Gamma_d\approx 3.72$ MeV, and $c_d\approx 56.79$
Gev$^{-3}$. 
For the quadrupole case we include both the IS-GQR and the IV-GQR.
The contributions of the GQRs to the photoabsorption
cross section due to the IS and IV $E2$ transitions are much smaller
than that from the GDR due to the $E1$ transition \cite{Roca}.
It renders difficult an accurate fit of the parameters of the GQRs from 
data on the photoabsoption cross section.
We use parameters of the IS-GQR   obtained in Ref.~\cite{IS1} 
from measurements of the IS $E2$ strength in inelastic scattering of
$\alpha$ particles at small angles: 
$\omega_q^{IS}\approx 10.89$ MeV and $\Gamma_q^{IS}\approx 3$ MeV.
We extracted the normalization constant from the energy weighted sum
rule (EWSR) (see, e.g., Refs.~\cite{Speth,Roca}) for the isoscalar quadrupole 
moment that to a good accuracy is exhausted by the IS-GQR \cite{IS1}. 
This gives $c_q^{IS}\approx 3984$ Gev$^{-5}$. 
For the IV-GQR we use parameters obtained in the recent most 
accurate measurement \cite{IV1}  via polarized Compton scattering:
$\omega_q^{IV}\approx 23$ MeV and $\Gamma_q^{IV}\approx 3.9$ MeV.
For the IV modes the EWSR may be violated by $\sim 20-30$\%  
due to the exchange potential in the nuclear Hamiltonian 
\cite{RMP63,Migdal_GDR,Roca}. The experimental data from 
Refs. \cite{IV1,IV19,IV20,IV11} give the exhaustion of the IV EWSR 
$100\pm 40$\%. We use the result of the most accurate measurement \cite{IV1}, 
that gives the exhaustion of the IV EWSR $56\pm 5$\%.
This leads to the normalization constant for the IV quadrupole 
polarizability $c_q^{IV}\approx 1524$ Gev$^{-5}$. 
The possible errors in the $c_q^{IV}$ 
are not very important because  anyway the IV contribution
turns out to be suppressed as compared to the IS one
due to bigger energy of the IV-GQR.

\begin{figure} [t]
\vspace{.7cm}
\begin{center}
\includegraphics[height=6.5cm]{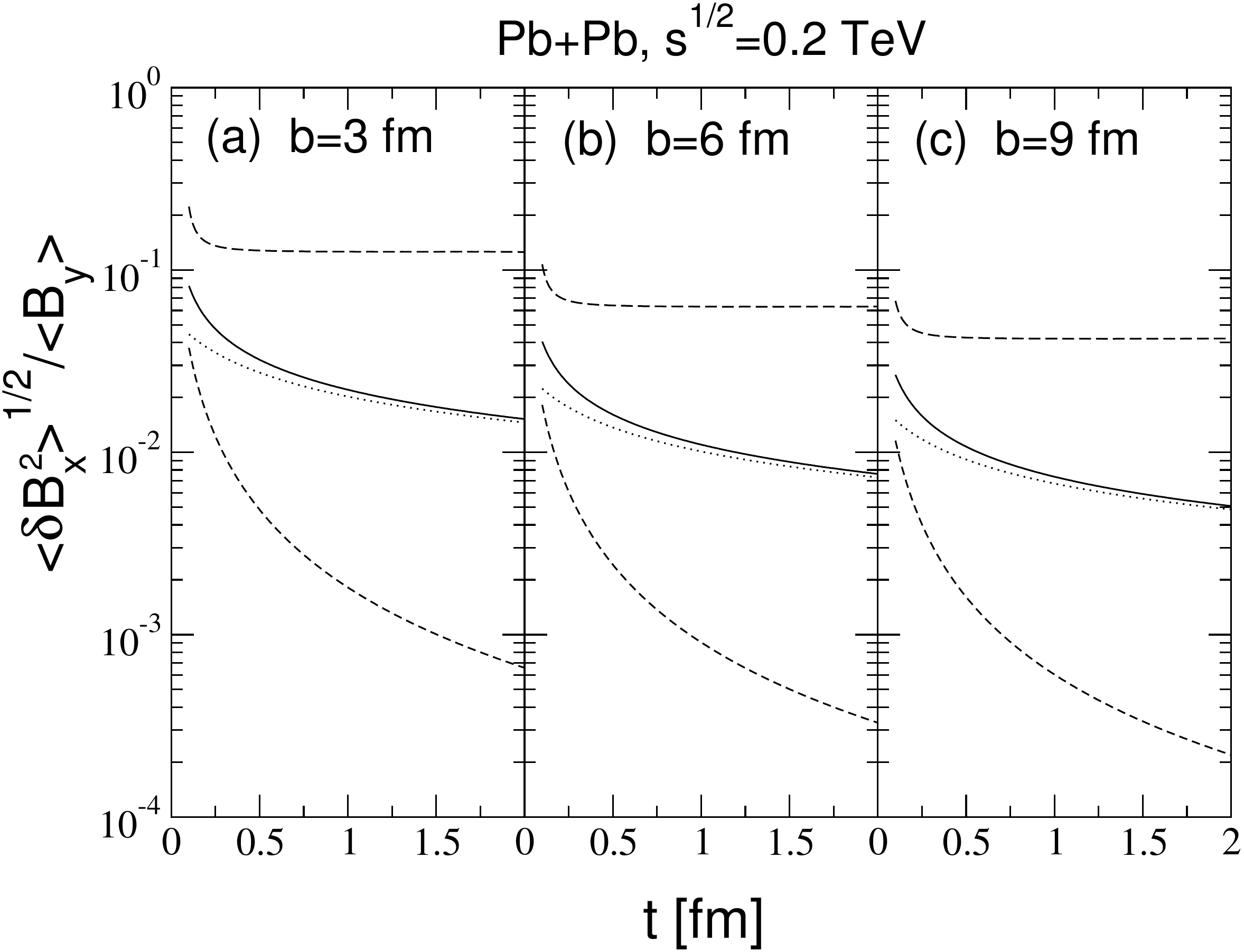}
\end{center}
\vspace{-0.5cm}
\caption[.]
{Ratio 
$\langle \delta B_x^2\rangle^{1/2}/\langle B_y\rangle$ versus $t$ 
at $\rb=0$ for Pb+Pb collisions at $\sqrt{s}=0.2$ TeV
for impact parameters $b=3$, $6$ and $9$ fm.
Results of quantum calculations: solid line is for the total contribution
of the GDR, IS-GQR and IV-GQR, dotted line is for the contribution of the GDR,
dashed line is for sum of the contributions of the IS-GQR and IV-GQR. 
Long-dashed lines show results of the classical Monte-Carlo calculation with
the WS nuclear density.}
\end{figure}
\begin{figure} [t]
\vspace{.7cm}
\begin{center}
\includegraphics[height=6.5cm]{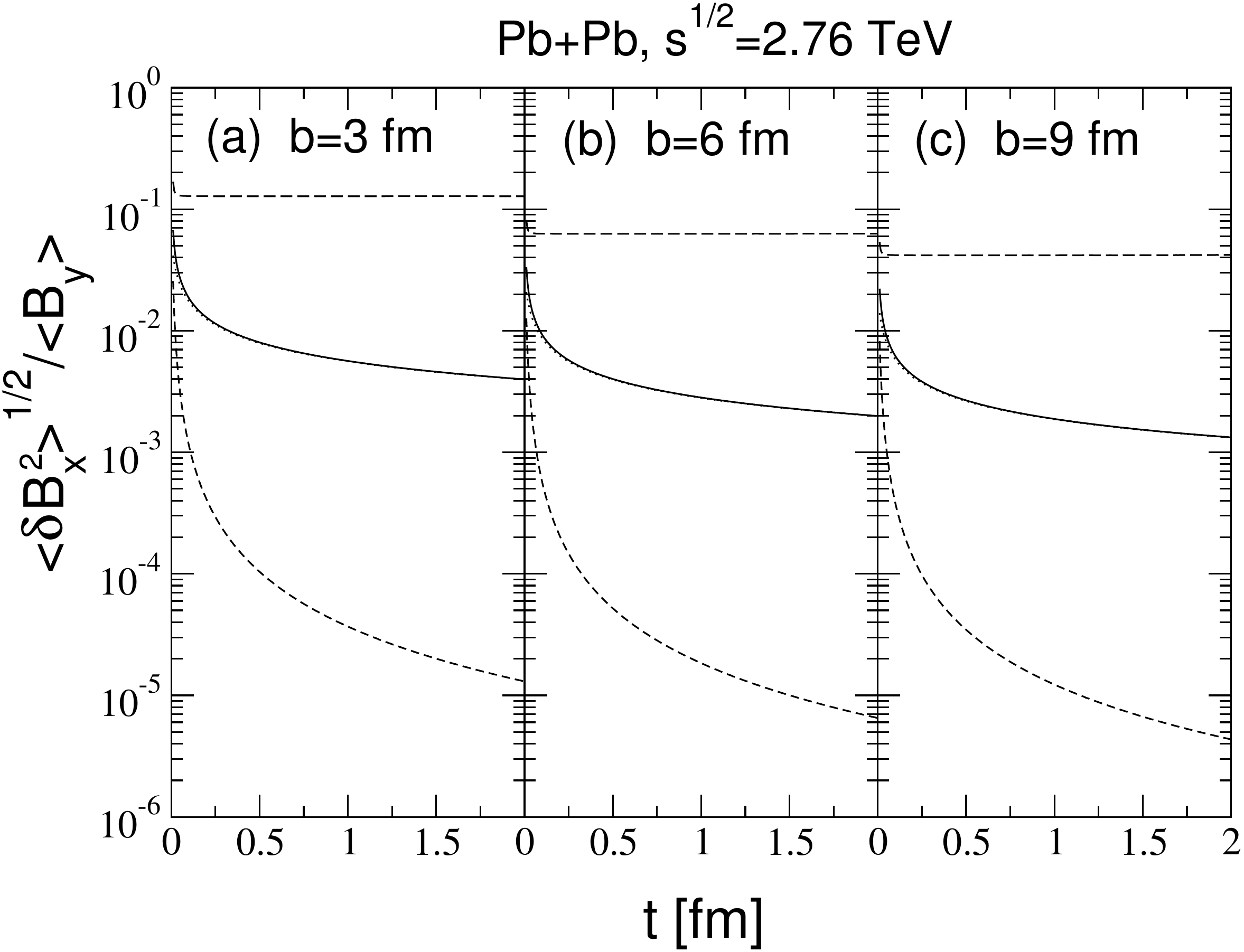}
\end{center}
\vspace{-0.5cm}
\caption[.]
{Same as in Fig.~2 but for $\sqrt{s}=2.76$ TeV.}
\end{figure}

\section{Numerical results} 
For applications the most interesting effect of the field fluctuation
is fluctuation of the direction of the magnetic field at the 
center of the plasma fireball. It is dominated
by the fluctuations of the component $B_x$ that vanishes without field 
fluctuations. In Figs.~2,~3 we show our 
results for $t$-dependence  
of the ratio $\langle \delta B_x^2\rangle^{1/2}/\langle B_y\rangle$
(which gives the typical angle between the magnetic field and the 
perpendicular to the reaction plane)
at $\rb=0$ for the impact parameters $b=3,$ $6$ and $9$ fm
for RHIC energy $\sqrt{s}=0.2$ and LHC energy $\sqrt{s}=2.76$ TeV.
For comparison we also show predictions of the classical Monte-Carlo
calculations for the WS nuclear density.
We present the results for $t_{min}<t<2$ fm with 
$t_{min}=0.1 (0.01)$ fm for $\sqrt{s}=0.2(2.76)$ TeV. This values of $t_{min}$,
in terms of the equations (\ref{eq:100}) and (\ref{eq:120}), 
correspond to the situation with 
$|\rb_{1,2}-\rb_{A}|/R_A\gsim 2-3$, 
when the observation point
is still not too close to the nuclei (in their rest frames). 
In this case
even at minimal $t$ 
the approximation of the point-like nuclei should still be reasonable.  
Figs.~2,~3 show that at $t\sim t_{min}$ in the quantum picture 
$\langle \delta B_x^2\rangle^{1/2}/\langle B_y\rangle$ 
is smaller than in the classical one by a factor of $\sim 3$.
One can see that at $t\sim t_{min}$ in the quantum picture  the quadrupole 
contribution is of the order of the dipole one. But the relative contribution 
of the GQRs falls steeply with increase of $t$. And it
becomes very small at $t\gsim 0.5(0.05)$ fm for $\sqrt{s}=0.2(2.76)$ TeV.
In this region the quantum picture gives  
$\langle \delta B_x^2\rangle^{1/2}/\langle B_y\rangle$
that is smaller than the classical prediction by a factor of $\sim 3-8$ 
for $\sqrt{s}=0.2$ TeV and by a factor of $\sim 6-30$ for for 
$\sqrt{s}=2.76$ TeV.
Note that the quadrupole contribution comes mostly from the IS mode
that has a smaller excitation energy. 
The field fluctuations also lead to nonzero values of the transverse
electric field $\langle \delta E_{x,y}^2\rangle^{1/2}$ at $\rb=0$,
that vanishes for the average field. 
The results for $\langle \delta E_{x,y}^2\rangle^{1/2}$ at $\rb=0$ 
are very similar to that for magnetic field.

Thus we see that in the quantum picture 
fluctuations of the direction of the magnetic field relative to the reaction
plane turns out to be considerably smaller than in the classical picture. 
The reduction of the field fluctuations in the quantum picture comes partly
from smaller fluctuations of the dipole and quadrupole moments and partly from
the dynamical quantum effects in the electromagnetic fields.
The latter lead to an increase of the difference between the quantum
and the classical models with increasing $t$. This quantum effects 
for the electromagnetic fields become important in the regime 
when $R=|\rb_{1,2}-\rb_A|$ in (\ref{eq:100}), (\ref{eq:120}) 
is large as compared to the inverse giant resonance excitation energies. 
The reduction of the fluctuations
of the dipole and quadrupole moments in the quantum picture can be
demonstrated by comparing the $\langle 0 |\db^2|0\rangle$ and
$\langle 0 |(Q_{ij}/6)^2|0\rangle$ with their classical
counterparts predicted by the Monte-Carlo calculations with the 
WS nuclear density.
In quantum picture the dipole moment squared can be
written as \cite{Z1} 
\beq
\langle 0 |\db^2|0\rangle=\frac{1}{\pi}
\int_{0}^{\infty}d\omega \mbox{Im}\alpha_d(\omega)\,.
\label{eq:210}
\eeq
It gives a value by a factor of $\sim 5$ smaller than the prediction
of the classical Monte-Carlo calculation with the WS nuclear density.
For the quadrupole mode one can easily obtain from (\ref{eq:130}), 
(\ref{eq:190})  a similar formula
\beq
\langle 0 |(Q_{ij}/6)^2|0\rangle=\frac{1}{\pi}
\int_{0}^{\infty}d\omega \mbox{Im}\alpha_q(\omega)\,.
\label{eq:220}
\eeq
Calculations 
using this formula show that the quantum result is smaller than 
prediction of the classical Monte-Carlo calculation with the 
WS nuclear density by a factor of $\sim 8$.
The fact the classical treatment based on the WS nuclear density overestimates 
the fluctuations of the dipole and quadrupole moments means 
that it overestimates the ellipsoidal fluctuations of the  nuclear density.
Note that this may be very important for the event-by-event hydrodynamic
simulations of $AA$ collision that presently ignore possible
collective effects in the nuclear distributions.

\section{Conclusion} 
In this work
we have performed a quantum
analysis of fluctuations of the electromagnetic field in $AA$ collisions
at RHIC and LHC energies.    
We use the FDT formalism of \cite{LL9} accounting for the contributions 
to the nucleus polarizability of the dipole and quadrupole modes.
We have found that for each nucleus the contribution of 
the IS and IV quadrupole modes
is of the order of that of the dipole mode when the distance between the
observation point and the center of the nucleus in the nucleus 
rest frame is $\sim 2R_A$. For the center of the QGP fireball
in the center mass frame of $AA$ collisions it corresponds
to time $t\sim 0.1(0.01)$ fm for RHIC(LHC) energy, and 
at $t\gsim 0.5(0.05)$ fm the dipole mode dominates the field fluctuations. 

Our quantum calculations show that effect of 
the field fluctuations is considerably smaller than that
in the classical Monte-Carlo simulation with the WS nuclear 
distribution. 
And in the quantum picture the fluctuations
of the direction of the magnetic field as compared to the mean field 
turn out to be very small.
Our results do not support a qualitative analysis \cite{Tuchin_quantum}, 
where it was argued that the quantum diffusion
of the protons may be very important. 

In the present analysis we have discussed the case of the spherical 
$^{208}$Pb nucleus. For collisions of the deformed $^{197}$Au nuclei,
that have been studied in RHIC experiments, one should account for a 
non-zero mean quadrupole moment. This can modify the contribution
of the quadrupole fluctuations. But the magnitude of the quantum quadrupole 
fluctuation around the equilibrium shape for the deformed nuclei is similar to
that for spherical ones (see, e.g., Refs.~\cite{Scamps1,Scamps2}).
Since the GDR peak in the photoabsorption cross section 
for $^{197}$Au \cite{GDR_Au}
is very similar to that for the $^{208}$Pb nucleus \cite{GDR_Pb}, 
the dominating contributions of the GDR to the field fluctuations
for these nuclei are also similar.
For this reason the conclusion that the classical approach
overestimates the field fluctuations should hold
for Au+Au collisions as well.

\begin{acknowledgments}
I am grateful to Sergey Kamerdzhiev  for helpful communications on physics
of the giant resonances.
This work is supported by the RScF grant 16-12-10151.
\end{acknowledgments}


\begin{thebibliography}{99}

\bibitem{Kharzeev_B1}
D.E. Kharzeev, L.D. McLerran, and H.J. Warringa,
Nucl. Phys. A{\bf 803}, 227 (2008)
[arXiv:0711.0950].

\bibitem{Toneev_B1}
V. Skokov, A.Yu. Illarionov, and V. Toneev,
Int.~J.~Mod.~Phys. A{\bf 24}, 5925 (2009)
[arXiv:0907.1396].

\bibitem{Tuchin_B}	
K. Tuchin,
Phys.~Rev. C{\bf 88}, 024911  (2013)
[arXiv:1305.5806].



\bibitem{Z_maxw}
B.G. Zakharov, Phys. Lett. B{\bf 737}, 262 (2014)
[arXiv:1404.5047].


\bibitem{Kharzeev_CME_rev} 	
D.E. Kharzeev,
Prog.~Part.~Nucl.~Phys. {\bf 75}, 133 (2014)
[arXiv:1312.3348].

\bibitem{T1}
K. Tuchin,
Phys. Rev. C{\bf 91}, 014902 (2015)
[arXiv:1406.5097].

\bibitem{Z_syn}
B.G. Zakharov, Eur.~Phys.J. C{\bf 76}, 609 (2016)
[arXiv:1609.04324].

\bibitem{HQ_dif1}
K. Fukushima, K. Hattori, and Y. Yin,
Phys.~Rev. D{\bf 93}, 074028  (2016)
[arXiv:1512.03689].

\bibitem{HQ_dif2}
S.I. Finazzo, R. Critelli, R. Rougemont, and J. Noronha,
Phys.~Rev. D{\bf 94}, 054020  (2016)
[Erratum-ibid.\ D{\bf 96}, 019903  (2017)]
[arXiv:1605.06061].


\bibitem{MHD1}
V. Roy, S. Pu, L. Rezzolla, and D.H. Rischke,
Phys.Rev. C{\bf 96}, 054909  (2017)
[arXiv:1706.05326].


\bibitem{MHD2}
A.~Das, S.S. Dave, P.S. Saumia, and A.M. Srivastava,
Phys.Rev. C{\bf 96}, 034902 (2017),
[arXiv:1703.08162].

\bibitem{MHD3}
V. Roy, Universe {\bf 3}, 82 (2017).

\bibitem{Skokov_B}
L. McLerran and V. Skokov,
Nucl. Phys. A{\bf 929}, 184 (2014)
[arXiv:1305.0774].



\bibitem{Liao_MC}
J. Bloczynski, X.-G. Huang, X. Zhang, and J. Liao,
Phys. Lett. B{\bf 718}, 1529 (2013)
[arXiv:1209.6594].



\bibitem{Skokov_MC} 	
A. Bzdak and V.  Skokov,
Phys.~Lett. B{\bf 710}, 171 (2012)
[arXiv:1111.1949].


\bibitem{Deng_MC}
W.-T. Deng and X.-G. Huang,
Phys. Rev. C{\bf 85}, 044907 (2012) 
[arXiv:1201.5108].

\bibitem{Roy_MC}
V. Roy and S. Pu,
Phys.Rev. C{\bf 92}, 064902  (2015)
[arXiv:1508.03761].

\bibitem{BM}
A. Bohr and B.R. Mottelson, 
{\it Nuclear structure}, vol. I \& II. W.A. Benjamin, Inc., 1975.



\bibitem{Greiner}
W. Greiner and J.A. Maruhn, 
{\it Nuclear models}, Berlin, Springer, 1996.

\bibitem{Speth}
S. Kamerdzhiev, J. Speth, and G. Tertychny,
Phys. Rept. {\bf 393}, 1 (2004) 
[nucl-th/0311058].


\bibitem{Roca}
X. Roca-Maza and N. Paar,
Prog.~Part.~Nucl.~Phys. {\bf 101}, 96 (2018)
[1804.06256].


\bibitem{Callen}
H.B. Callen and T.A. Welton, Phys. Rev. {\bf 83}, 34 (1951).

\bibitem{LL9} E.M. Lifshits and L.P. Pitaevski,
{\it Statistical Physics, Part 2 (Landau Course of 
Theoretical Physics Vol. 9)}, Oxford, Pergamon
  Press, 1980. 

\bibitem{Z1}
B.G. Zakharov,
JETP Lett. {\bf 105}, 758 (2017) 
[arXiv:1703.04271].

\bibitem{LL4} V.B. Berestetski, E.M. Lifshits, and L.P. Pitaevski,
{\it Quantum Electrodynamics (Landau Course of 
Theoretical Physics Vol. 4)}, Oxford, Pergamon
  Press, 1979. 


\bibitem{Migdal_GDR}
A.B. Migdal, A.A. Lushnikov, and D.F. Zaretsky,
Nucl. Phys. {\bf 66}, 193 (1965).





\bibitem {LL2} L.D.~Landau and E.M.~Lifshitz,  
{\it Classical Theory of Fields},   
Butterworth-Heinemann, Oxford,
UK, 1987.


\bibitem{Greiner_PR151}
M. Danos, W. Greiner, and C.B. Kohr,
Phys. Rev. {\bf 151}, 761  (1966). 

\bibitem{GDR_Pb}
A. Tamii {\it et al.},
Phys. Rev. Lett. {\bf 107}, 062502 (2011)
[arXiv:1104.5431].

\bibitem{IS1}
D.H. Youngblood, Y.W. Lui, H.L. Clark, B. John, Y. Tokimoto, and X. Chen 
Phys.~Rev. C{\bf 69}, 034315  (2004). 


\bibitem{IV1}
S.S. Henshaw, M.W. Ahmed, G. Feldman, A.M. Nathan, and H.R. Weller,
Phys.~Rev.~Lett. {\bf 107}, 222501  (2011).

\bibitem{RMP63}
E. Hayward,
Rev.~Mod.~Phys. {\bf 35}, 324 (1963). 


\bibitem{IV19}
R. Leicht, M. Hammen, K.P. Schelhaas, and B. Ziegler,
Nucl.~Phys. A{\bf 362}, 111 (1981).

\bibitem{IV20}
K.P. Schelhaas, J.M. Henneberg, M. Sanzone-Arenhövel, N. Wieloch-Laufenberg,
U. Zurmühl, B. Ziegler, M. Schumacher, and F. Wolf. 
Nucl.~Phys. A{\bf 489}, 189 (1988).
 
\bibitem{IV11}
D.S. Dale, R.M. Laszewski, and R. Alarcon,
Phys.~Rev.~Lett. {\bf 68},  3507(1992).


\bibitem{Tuchin_quantum}
R. Holliday, R. McCarty, B. Peroutka, and K. Tuchin,
Nucl. Phys. A{\bf 957}, 406 (2017)
[arXiv:1604.04572].



\bibitem{Scamps1}
G. Scamps and D. Lacroix,
Phys.~Rev. C{\bf 88}, 044310  (2013)
[arXiv:1307.1909].

\bibitem{Scamps2}
G. Scamps and D. Lacroix,
Phys.~Rev. C{\bf 89}, 034314  (2014) 
[arXiv:1401.5211].

\bibitem{GDR_Au} A. Veyssiere, H. Beil, R. Bergere, P. Carlos, and
  A. Lepretre, Nucl. Phys. A{\bf 159}, 561 (1970).


\end{thebibliography}
\end{document}